\documentclass[twocolumn,showpacs,aps,floatfix]{revtex4}
\usepackage{graphicx}
\begin{document}

\newcommand{\D}{\displaystyle} 
\newcommand{\T}{\textstyle} 
\newcommand{\SC}{\scriptstyle} 
\newcommand{\SSC}{\scriptscriptstyle} 

\def\AJ{{\it Astron. J.} }
\def\ARAA{{\it Annual Rev. of Astron. \& Astrophys.} }
\def\ApJ{{\it Astrophys. J.} }
\def\ApJL{{\it Astrophys. J. Letters} }
\def\ApJS{{\it Astrophys. J. Suppl.} }
\def\ApP{{\it Astropart. Phys.} }
\def\AA{{\it Astron. \& Astroph.} }
\def\AAR{{\it Astron. \& Astroph. Rev.} }
\def\AAL{{\it Astron. \& Astroph. Letters} }
\def\AASu{{\it Astron. \& Astroph. Suppl.} }
\def\AN{{\it Astron. Nachr.} }
\def\IJMP{{\it Int. J. of Mod. Phys.} }
\def\JGR{{\it Journ. of Geophys. Res.} }
\def\JHEP{{\it Journ. of High En. Phys.} }
\def\JPhG{{\it Journ. of Physics} {\bf G} }
\def\MNRAS{{\it Month. Not. Roy. Astr. Soc.} }
\def\Nature{{\it Nature} }
\def\NewAR{{\it New Astron. Rev.} }
\def\PASP{{\it Publ. Astron. Soc. Pac.} }
\def\PhFl{{\it Phys. of Fluids} }
\def\PLB{{\it Phys. Lett.}{\bf B} }
\def\PR{{\it Phys. Rev.} }
\def\PRD{{\it Phys. Rev.} {\bf D} }
\def\PRL{{\it Phys. Rev. Letters} }
\def\RMP{{\it Rev. Mod. Phys.} }
\def\Science{{\it Science} }
\def\ZfA{{\it Zeitschr. f{\"u}r Astrophys.} }
\def\ZfN{{\it Zeitschr. f{\"u}r Naturforsch.} }
\def\etal{{\it et al.}}

\hyphenation{mono-chro-matic sour-ces Wein-berg
chang-es Strah-lung dis-tri-bu-tion com-po-si-tion elec-tro-mag-ne-tic
ex-tra-galactic ap-prox-i-ma-tion nu-cle-o-syn-the-sis re-spec-tive-ly
su-per-nova su-per-novae su-per-nova-shocks con-vec-tive down-wards
es-ti-ma-ted frag-ments grav-i-ta-tion-al-ly el-e-ments me-di-um
ob-ser-va-tions tur-bul-ence sec-ond-ary in-ter-action
in-ter-stellar spall-ation ar-gu-ment de-pen-dence sig-nif-i-cant-ly
in-flu-enc-ed par-ti-cle sim-plic-i-ty nu-cle-ar smash-es iso-topes
in-ject-ed in-di-vid-u-al nor-mal-iza-tion lon-ger con-stant
sta-tion-ary sta-tion-ar-i-ty spec-trum pro-por-tion-al cos-mic
re-turn ob-ser-va-tion-al es-ti-mate switch-over grav-i-ta-tion-al
super-galactic com-po-nent com-po-nents prob-a-bly cos-mo-log-ical-ly
Kron-berg Berk-huij-sen}
\def\simle{\lower 2pt \hbox {$\buildrel < \over {\scriptstyle \sim }$}}
\def\simge{\lower 2pt \hbox {$\buildrel > \over {\scriptstyle \sim }$}}
\def\intunits{{\rm s}^{-1}\,{\rm sr}^{-1} {\rm cm}^{-2}}

\title{Cosmic ray electrons and positrons from supernova explosions of
    massive stars}

\author{P.L.~Biermann,}\vspace*{-30pt} 
\affiliation{MPI for Radioastronomy, Bonn, Germany}
\altaffiliation[Also at ] {Dept. of Phys. \& Astr., Univ. of Alabama,
Tuscaloosa, AL, USA}
\altaffiliation{Dept. of Phys., Univ. of Alabama at Huntsville, AL, USA}
\altaffiliation{FZ Karlsruhe, and Phys. Dept., Univ. Karlsruhe, Germany}
\author{J.K.~Becker} \vspace*{-20pt}
\affiliation{Institution f{\"o}r Fysik, G{\"o}teborgs Univ., Sweden}
\altaffiliation[Also at ] {Dept.\ of Phys.\& Astron., Ruhr-Univ.\ Bochum,
  Germany}
\author{A.~Meli,}\vspace*{-30pt}
\affiliation{ECAP, Physik. Inst. Friedrich-Alexander
 Univ. Erlangen-N{\"u}rnberg, Germany}
\author{W.~Rhode,}\vspace*{-30pt}
\affiliation{Dept. of Phys., Univ. Dortmund, Dortmund, Germany}
\author{E.-S.~Seo,}\vspace*{-30pt}
\affiliation{IPST and Dept. of Physics, Univ. of Maryland, College Park,
 MD, USA}
\author{T.~Stanev}\vspace*{-30pt}
\affiliation{Bartol Research Inst. and Dept. of Phys. and Astronomy,
 Univ. of Delaware, Newark, DE, USA}

\begin{abstract}
We attribute the recently discovered cosmic ray electron and cosmic ray 
    positron excess components and their cutoffs to the acceleration
    in the supernova 
shock in the polar cap of exploding Wolf Rayet and Red Super Giant 
stars. Considering a spherical surface at some radius around such a 
star, the magnetic field is radial in the polar cap as opposed to most 
of $4 \pi$ (the full solid angle), 
where the magnetic field is nearly tangential. 
This difference yields a flatter spectrum, and also an enhanced positron 
injection for the cosmic rays accelerated in the polar cap.
This reasoning naturally explains the observations. Precise spectral 
measurements will be the test, as this predicts a simple $E^{-2}$ 
spectrum for the new components in the source, steepened to $E^{-3}$ in 
observations with an $E^{-4}$ cutoff.
\end{abstract}
\maketitle

 Recently an excess of both cosmic ray (CR) positrons and cosmic ray 
 electrons has been detected by three instruments, PAMELA~\cite{adrietal08}, 
 ATIC~\cite{changetal08}, and H.E.S.S.~\cite{aharetal08}. 
 The ATIC and H.E.S.S. results on cosmic ray electrons are consistent
 with a discovery of an excess, compared to the normal measured
 spectrum of $E^{-3.26 \pm 0.06}$~\cite{WSaB99}; we emphasize that
 also the H.E.S.S. result is above such an extrapolation.
 Both excesses take the form of a 
 flatter component emerging from below a steeper, perhaps normal, 
 component; this has been interpreted in many ways, such as, e.g., the 
 decay of a new particle representing dark matter~\cite{bergsetal08},
 or as evidence of a nearby special source~\cite{yukseletal08}.
 Here we wish to point out that such spectral components 
 are expected from particle acceleration in the explosion of stars with 
 magnetic winds~\cite{abbottetal84}: There is a small polar cap 
 component, where the acceleration in a supernova (SN) shock proceeds 
 with the magnetic field parallel to the shock normal, yielding an 
 $E^{-2}$ spectrum~\cite{drury83}. At the same time, over most of $4 \pi$ 
 the magnetic field is best approximated by an Archimedian
 spiral~\cite{parker58}, implying a near-perpendicular configuration
 for acceleration~\cite{jokip87}. 
 In such a situation the curvature becomes important, and 
 the acceleration gives a slightly steeper spectrum and is faster - once 
 injection has been effected. This was discussed again recently by Meli 
 \& Biermann~\cite{AMaB06}. The spectrum predicted is
 $E^{-7/3 - 0.02 \pm 0.02}$~\cite{plb93,plbcas93,plb06}. The normal 
 observed cosmic ray electron spectrum for $E \, > \, 10$ GeV provides a 
 test, since it gives $E^{-2.26 \pm 0.06}$ after correcting for
 losses~\cite{kard62,WSaB99}.

 Our approach here is to adopt the following point of
 view~\cite{plb93,plbetal01}:\\ 
 a) Most of the interactions of cosmic rays happen near the sources,
 and the escape from the Galaxy is governed by a simple Kolmogorov
 description~\cite{rick77}.\\
 b) The bend in the spectrum, the knee at $10^{15}$ eV is due to
 spatial limitations given by a shocked shell racing through a
 Parker-type magnetic wind~\cite{plb93,plb06}.\\
 c) We do invoke the physics of the stars that explode~\cite{hegeretal03}, 
 and distinguish three zero age mass ranges of massive stars which
 explode: The stars between about 8 and 15 solar masses, which explode
 into the interstellar medium; the stars between 15 and about 25 solar
 masses which explode as Red Super Giant (RSG) stars, 
 and the stars above about 25 solar masses, which explode as Blue Super 
 Giant, or Wolf Rayet (WR) stars. Both RSG stars and WR stars explode 
 into their stellar wind~\cite{volkbier88,plbcas93}, which is magnetic,
 and enriched from exposing the deeper layers of the star through mass
 ejections~\cite{langheg99}.

 It was shown in Ref.~\cite{stanevetal93} that such a combination of two 
 components, a dominant $E^{-7/3}$ spectrum, with an additional $E^{-2}$, 
 injected at the level of a few percent, yields a good fit to the cosmic 
 ray air shower data.
   Protons and heavier nuclei spectra are well described right
   through the knee of the cosmic ray spectrum (near $10^{15}$ eV)
   after accounting for transport that steepens the
   spectra by $E^{-1/3}$. 
   At this energy the polar cap component increases the overall
   flux by about a factor of 2, before cutting off due to spatial
   constraints~\cite{amh84,plb93}.
   The theoretical assumptions about acceleration in highly
   oblique shocks~\cite{jokip87,AMaB06} used in Ref.~\cite{plb93}
   were justified by the good agreement with data~\cite{WSMB98}:
   prediction $E^{-8/3}$, data $E^{-2.68 \pm 0.02}$. 
   This earlier success encouraged us to apply precisely 
   the same concept here.
 Additional enhancements of magnetic fields may 
 occur also in such a situation~\cite{belluc01}.

 We find, that using just the parameters of earlier papers it is possible 
 to explain both the cosmic ray electron spectrum and the cosmic ray
 positron excess.

 At GeV energies most of the cosmic ray (CR) electrons are accelerated in 
 supernova (SN) shocks, running through the interstellar medium (ISM), 
 the ISM-SN CRs. This predicts a spectrum of
 $E^{-2.42 \pm 0.04}$~\cite{plbstrom93}, 
 in agreement with radio data of other 
 galaxies, for which the leakage energy dependence modifies the predicted 
 spectrum to $E^{-2.75 \pm 0.04}$~\cite{plb97}. Already the 
 data beyond about 10 GeV suggest that the wind-supernovae
 cosmic rays may have taken 
 over also for cosmic ray electrons~\cite{WSaB99}.

 Helium, Carbon and heavier nuclei give an indication~\cite{stanevetal93}
 of what fraction of $4 \pi$ the polar cap component may have.
 This component   reaches the same flux by itself as the rest
 of $4 \pi$ at $3 \, 10^{6}$ GeV/nucleus for CNO, yielding a
 surface fraction of about 2 percent.
     Since this fraction does
     not depend on distance from the star (see below) we assume that
     it defines the energy when the polar cap spectrum exceeds the
     $E^{-7/3}$ component.

 Electrons, however, are injected at about 30 MeV, the lowest energy at 
 which they see the shock~\cite{prothplb96}. This 
 number derives from the injection condition for electrons,
 that they must ``see" the waves excited by the ions freshly injected
 by shocks in the assumption, that the plasma
 is dominated by ionized Hydrogen, and that the shock velocity is about 
 10,000 km/s. In a Wolf-Rayet star wind the main element is, however, not 
 Hydrogen, but heavier nuclei, and already before the star 
 explodes as a supernova, there are accelerated electrons: The velocity 
 of the shocks caused by instabilities in the radiation driving is of 
 order $\simge \, 1000$ km/s~\cite{owocetal88}, and so the 
 electron energy at injection then is at about $\simge \, 6$ MeV.

 This immediately implies that the polar cap component of cosmic ray 
 electrons should rise to a flux equal of the rest at $\simge \, 400$ 
 GeV, matching the uncertain observed energy of about 300 - 500 GeV. So 
 at this energy the sum of the two components is twice the base spectral 
 component.
 We interpret the ATIC data here as a $E^{-3}$ component, rising above 
 the base spectral component of $E^{-10/3}$ around 30 to 100 GeV.

 Cosmic ray positrons derive from collisions of nuclei, and formation of 
 nuclei to the left of the valley of stability which decay in 
 $\beta^{+}$-emission, and also from pion production and decay. However, 
 here we have to remember, that acceleration is faster for perpendicular 
 shocks, by a factor up to $c/(3 V_{sh})$, 
 probably more like 2 - 3~\cite{AMaB06}. This 
 implies that the polar cap component is more efficient in producing 
 positrons because of its slower acceleration and higher interaction
 probability. Since the hadronic interaction cross section is almost
    energy independent and does not introduce a break in the spectrum, 
    the polar cap component becomes dominant 
    at an energy between $2^3$ to  $10^{3}$ lower than for electrons,
 i.e. between  0.5 to 60 GeV. 30 GeV seems to be 
 compatible with the data, suggesting that the enhancement given by 
 perpendicular shock acceleration is about a factor of 2 - 3. However, as 
 there is a second source of positrons at lower energy, resulting from 
 interaction in the immediate environment of massive exploding stars and
 in the interstellar space~\cite{rjp82,moskstr98}, the cross-over may be
 at lower energy, suggesting a possibly 
 higher efficiency enhancement. In a CR-positron to 
 CR-(electron+positron) ratio this results in a rise with $E^{+1/3}$.

  From all these interaction sites here should be a corresponding 
 neutrino-emission with a spectrum of $E^{-2}$. On the other hand, as the 
 cosmic ray electrons approach a spectrum of $E^{-3}$ themselves, in the 
 ratio positrons to electrons we approach a constant from somewhere in 
 the range 30 - 100 GeV, when both electron and positron components are 
 dominated by the polar cap.

 Here we discuss the second positron component at low energy, introduced 
 above, which distorts the positron spectrum:

 The wave-field in the magnetic field excited by the cosmic rays of 
 spectrum $E^{-7/3}$~\cite{plbcas93} in the predecessor 
 stellar wind naturally yields a specific spectrum of turbulence, in 
 energy per volume and wave number $I(k) \, \sim \, k^{-13/9}$, where $k 
 \, = \, 2 \pi /r_g$, and $r_g \, = \, p c/(Z e B)$, the Larmor radius 
 (here $p$ is the momentum of the particle, $Z$ its charge, $c$ the speed 
 of light, $e$ the elementary charge, and $B$ the ambient magnetic field 
 component perpendicular to the motion of the particle). This spectrum of 
 magnetic irregularities then governs the transport of cosmic rays and 
 the cosmic ray interaction as a function of energy. This in turn gives 
 rise to a secondary to primary ratio going as $E^{-5/9}$~\cite{plb97}.
 This prediction was confirmed in Ref.~\cite{ptus99} which showed 
 that the best fit of the secondary to primary ratio had an energy 
 dependence of $E^{-0.54}$. However, there is also another spectral 
 component of turbulence induced by instabilities leading to many weak 
 shock waves $I(k) \, \sim \, k^{-2}$, which is steeper. The total summed 
 spectrum of turbulence has then a cross-over towards lower wave-numbers, 
 corresponding to higher particle energies; this spectrum $I(k) \, \sim 
 \, k^{-2}$ induces no energy dependence of the production of 
 secondaries. For the most massive stars, exploding as WR stars, we 
 estimate the cross-over to correspond to somewhere near 10 GeV in 
 electron energy. For somewhat lower mass stars, those exploding as red 
 super giant stars, we argue, that there the cross-over between the two 
 spectral regimes of turbulence is at lower energies, or higher 
 wavenumbers, since the winds are less powerful. The cosmic ray induced 
 turbulence is driven by the mass flow through the supernova shock, and 
 so a wind of lesser density produces a weaker cosmic ray induced wave 
 field. This results in secondaries having the same spectrum as the 
 primaries, for WR stars above about 10 GeV, and for RSG stars at much 
 lower energy, disregarding for a moment the polar cap component.

 This helps understand the diffuse gamma-ray emission of the disk of our 
 Galaxy~\cite{casaetal04}, as interaction near the RSG stars, more 
 abundant than the WR stars. This model interprets all the secondary to 
 primary ratios at low energy~\cite{ptus99}.

 This may then explain the low energy cosmic ray positrons, seen with 
 PAMELA~\cite{adrietal08}, as argued earlier.

  The acceleration time for cosmic ray particles in strong shocks
  is~\cite{drury83,jokip87} $\tau_{acc} \; = \; {8 \, 
  \kappa}/{V_{sh}^{2}}$. Comparison of $\tau_{acc}$ to the synchrotron loss
  time $\tau_{syn} \; = \; {6 \pi m_e c}/{\sigma_T \gamma_e B^{2}}$ gives a 
  limit on the maximal energy of the electrons or positrons.

  In the polar cap $B \, \sim \, r^{-2}$, while over most of $4 \, 
  \pi$ $B \, \sim \, r^{-1}$. In the polar cap we assume maximal
  turbulence and Bohm diffusion $\kappa \, = \, (1/3) r_L c$.
  In the rest of the surface we adopt the
  approximation~\cite{jokip87} of $\kappa \, = \, r_L \, V_{sh}$.
  The leads to maximum $\gamma_{e, max} \; = 
  \; 1.5 \cdot 10^{6} \, B_{0.5, 14}^{-1/2} \, V_{sh, 9} \, 
  \left({r}/{r_0}\right)$ for the polar cap case, where $B_{0.5, 
  14}^{-1/2}$ is the magnetic field at radius $10^{14}$ cm in 
  units of 3 Gau{\ss}, and $V_{sh, 9}$ is the SN shock velocity
  in units of $10^{9}$ cm/s. 
  For most of $4 \pi$ the maximum is $\gamma_{e, max} \; = \; 4.5 \cdot 
  10^{6} \, B_{0.5, 14}^{-1/2} \, V_{sh, 9}^{1/2} \, 
  \left({r}/{r_0}\right)^{1/2}$.
  Using the adopted values of the magnetic field and shock velocity, at 
  $10^{16}$ cm radial distance covered by the SN-shock in the wind the 
  maximal energy of the CR electrons from the polar cap will dominate.
  Since the expressions used for $\kappa$ for both segments of the
  star's surface  are simplified both energies are likely to be smaller.
  But $\gamma_{max}$ from the polar cap increases as $r$ while for most
  of $4 \pi$ it increases as $\sqrt{r}$~\cite{plb97}.
  This implies that the polar cap component will quickly pass the other 
  component in maximal energy.

  Last, we check whether the angular fraction is really distance 
  independent, as required by such a model. Combining the results
  derived in Refs.~\cite{plb93,plbcas93,jokip87,volkbier88,AMaB06}
  we match the acceleration time scale in the polar cap with the 
  acceleration time over most of $4 \pi$.
  In the limit of a small angular extent $\theta$ of 
  the polar cap this gives: $\theta \; = \; 3 \, ({B_{0, p.c.}}/{B_{0, 
  4 \pi}}) \, ({V_{sh}}/{c})$ suggesting for ${V_{sh}}/{c} \, \simeq 
  \, 0.03$ a ratio of surface magnetic fields of
  ${B_{0, p.c.}}/{B_{0, 4 \pi}} \, \simeq \ 0.6$,  
  very close to unity and independent of radius. 
  It is an interesting question if for some stars 
  this ratio might be different.
  Accelerated electrons have energy loss given by the 
  synchrotron loss time $\tau_{syn}$ and they can also
  escape the Galaxy with $\tau_{leak} \; \sim 
  \; {\gamma_e}^{-1/3}$. Then we have 
  $N(E) \; \sim \; E^{-7/3 -1}$ spectrum in the limit
  $\tau_{syn} < \tau_{leak}$ and $N(E) \; \sim \; 
  E^{-7/3 -1/3}$, in the opposite limit. We can estimate from 
 observations, radio measurements as well as direct data, that the 
 switch-over is near 10 - 20 GeV.

  CR electrons reach us in a random walk, in
  which the distance actually travelled is given by
  $r \; \simeq \Delta r \, \sqrt{N}$, where $\Delta r (E)$ is the 
  diffusion scattering mean free path, the step, and $N$ is the number of 
  uncorrelated steps.
     The time this takes is given by
     $t \; \simeq ({\Delta r}/{c})\, N$, while
     $\Delta r \; \sim \; E^{1/3}$ and the timescale $t$ is
     $\sim \,1/E$. Therefore  $r \; \sim E^{-1/3}$.
     The region from which we obtain CR electrons is the volume for which 
     this time is less than the synchrotron time $\tau_{syn}$. 
     The source volume is proportional to  $r^{3} \; \sim E^{-1}$.
     Adopting the viewpoint that at high energies 
     the polar cap component of the CR contribution of the the wind-SNe 
     dominates implies then a combined spectrum of
     $\simeq E^{- 3 -1}$. So the CR electron spectrum is predicted in this 
     approximation to be $E^{-10/3}$, then $E^{-3}$, and thereafter $E^{-4}$
     which is consistent with the H.E.S.S. data~\cite{aharetal08}.
     Nearby massive stars exist~\cite{EAC09} and the associated 
     supernovae~\cite{PTPCT05} may have provided a large fraction
     of the cosmic rays we observe.
 A test for our predictions 
 would be a measurement of the exact spectrum of both 
 cosmic ray electrons as well as cosmic ray positrons.
 This model simply  predicts that their spectrum is a
 simple additional $E^{-2}$ component  at source.

The positron fraction should reach a plateau, the exact number depending 
on the dominant path to produce positrons, decay from isotopes pushed by 
photo-dissociation and spallation off the valley of stability in a $N,Z$ 
plot, or just simply pion decay.

  Ion collisions produce very few if any anti-protons and 
  contributions from WR star explosions will be negligible. 
  The RSG explosions will produce anti-protons~\cite{sinaetal01},
  and their polar cap contribution should come up at some higher
  energy, perhaps above 30 - 100 GeV.

Obviously, both cosmic ray electrons as well cosmic ray positrons are in 
their respective loss limit~\cite{kard62} at such energies, above 
about 10 - 20 GeV, where losses overpower diffusion and steepen the 
spectrum by unity; the limited spatial reach gives a further steepening 
by $E^{-1}$. Therefore there must be some yet higher energy where the 
losses from the nearest most recent source cut everything completely 
off; the ATIC and H.E.S.S. data suggest that this happens beyond 
energies of several TeV. It would be very interesting to measure the 
positron component to this energy.

We have proposed a simple explanation for the cosmic ray electron and 
cosmic ray positron components, in terms of the magnetic field topology 
in a magnetic wind~\cite{parker58}, and SN-induced shock acceleration in 
such a topology~\cite{jokip87,AMaB06}. The new component 
is just that population of energetic particles accelerated in the polar 
cap of massive magnetic stars with winds, when they explode.
\begin{figure}[thb]
\centering{\includegraphics[width=75truemm]{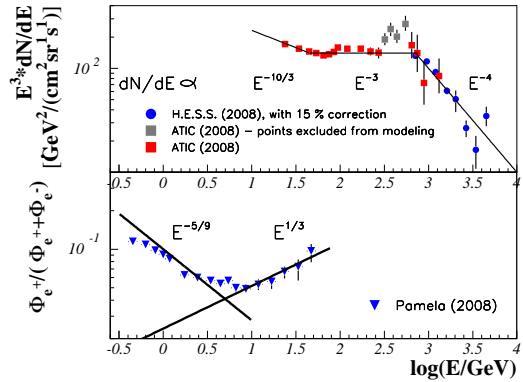}}
\caption{Model prediction with ATIC and H.E.S.S data for cosmic ray 
electrons: $E^{-10/3}$, $E^{-3}$, and $E^{-4}$ (upper panel), and for
PAMELA data on the positron/electron ratio: $E^{-5/9}$ and
$E^{1/3}$(lower panel). See Ref.~\cite{CE04} for charge dependent
solar modulation arguments.}
\label{plot}
\vspace*{-10pt}
\end{figure}
We attempted to explain a) the low energy PAMELA data, b) the higher 
energy PAMELA data, c) the ATIC data, and d) the H.E.S.S. data, all in 
the context of a basic picture proposed and worked out
    earlier~\cite{plb93,stanevetal93}, and 
consistent with other measurements.

In the cosmic rays produced by a shock running through such a wind there 
is always a polar cap component, of a few percent strength at injection, 
but with a flatter spectrum.

The explosions of Wolf Rayet stars and their cousins, the Red Super 
Giant stars, into their respective magnetic wind, may play a key role in 
allowing us to understand the physics of cosmic rays.

{\bf Acknowledgments}
Discussions with B. Harms, S. Casanova, L. Clavelli, C. 
Escobar, F. Halzen, A. Karle, U. Katz, M. Roth, V. de Souza, Ch. 
Stegmann, F. Tabatabaei, L. Trache, and St. Westerhoff are gratefully 
acknowledged. We would like to thank especially Ramin Sina for much 
effort on the anti-proton spectra resulting from this approach, and 
Sabrina Casanova for corresponding work for the $\gamma$-ray emission. 
JKB and WR would like to thank the MAGIC-collaboration for discussions.
 Support for work of PLB has come from the AUGER membership and
 theory grant 05 CU 5PD 1/2 via DESY/BMBF and VIHKOS.
 Support for JKB comes from the DFG grant BE-3714/3-1.
 WR is supported by the IceCube grant
 BMBF (05 A08PE1). Support for ESS comes from NASA grant NNX09AC14G and
 for TS comes from DOE grant UD-FG02-91ER40626.

\end{document}